\documentclass{aa}
\usepackage{psfig, graphics}
\usepackage{txfonts}
\usepackage[dvips]{graphicx}
\usepackage[authoryear]{natbib}
\usepackage[latin1]{inputenc}

\title{Mesoscale dynamics on the Sun's surface\\ from HINODE observations}

\author{Th.~Roudier\inst{1}, M.~Rieutord,\inst{1,2}, D.~Brito\inst{1,3},
F.~Rincon\inst{2}, J.M.~Malherbe\inst{4}, N.~Meunier\inst{5},
T.~Berger\inst{6}, Z.~Frank\inst{6}}

\date{Received \today  / Submitted }

\offprints{Th. Roudier}
 
\institute{
Laboratoire d'Astrophysique de Toulouse-Tarbes, Université de Toulouse, CNRS, 
57 Avenue d'Azereix, 65000 Tarbes, France
\and Laboratoire d'Astrophysique de Toulouse-Tarbes, Université de Toulouse, CNRS,
14 Avenue Edouard Belin, 31400 Toulouse, France
\and Laboratoire de Modélisation et d'Imagerie en Géosciences, Université de Pau et 
des Pays de l'Adour, CNRS, Avenue de l'Université, 64013 Pau Cedex, France 
\and LESIA, Observatoire de Paris, Section de Meudon, 92195 Meudon, France
\and LAOG, CNRS, Université Joseph Fourier, BP 43, 38041 Grenoble Cedex, France
\and Lockheed Martin Advance Technology Center, Palo Alto, CA, USA
}

\def\apjl{ApJ}%
\def\apjs{ApJS}%
\def\ao{Appl.~Opt.}%
\def\aap{A\&A}%
\begin{document}

\authorrunning{Roudier et al.}
\titlerunning{Supergranulation and  TFGs evolution from  Hinode observation}

\abstract{
}{
The interactions of velocity scales on the Sun's surface, from 
granulation to supergranulation are still not
understood, nor are their interaction with magnetic fields. We thus aim
at giving a better description of dynamics in the mesoscale range which
lies between the two scales mentioned above.
}{
We analyse a 48h high-resolution time sequence of the 
quiet Sun photosphere at the disk center obtained with the Solar Optical Telescope 
onboard Hinode. The observations, which have a field of view of 100~\arcsec$\times$ 
100~\arcsec, typically contain four supergranules. We monitor in detail the motion and 
evolution of granules as well as those of the radial magnetic field.
}{
This analysis allows us to better characterize Trees of Fragmenting Granules issued
from repeated fragmentation of granules, especially their lifetime statistics. Using 
floating corks advected by measured velocity fields, we show their crucial role 
in the advection of the magnetic field and in the build up of the network. Finally, 
thanks to the long duration of the time series, we estimate that the turbulent 
diffusion coefficient induced by horizontal motion is approximately 
$430\,\mathrm{km}^2\,\mathrm{s}^{-1}$.
}{
These results demonstrate that the long living families contribute
to the formation of the magnetic network  and suggest that supergranulation 
could be an emergent length scale building up as small magnetic elements are 
advected and concentrated by TFG flows. Our estimate for the 
magnetic diffusion associated with this horizontal motion might provide
a useful input for mean-field dynamo models.
}

\keywords{The Sun: Atmosphere -- The Sun: Granulation -- The Sun: Magnetic fields}
\maketitle

\section{Introduction}

The transport and evolution of magnetic flux over the solar surface 
are linked to the turbulent flows developing within the solar 
convection zone. Indeed, magnetic fields are observed in complex and 
hierarchical structures covering widely different scales which
emerge and disappear on the time-scales of turbulent convective patterns.
This continual reconfiguration of surface magnetism by surface flows
plays a significant role in determining the topology and evolution 
of coronal magnetic fields. It may notably influence the
triggering of eruptive events such as flares or sudden disappearance of 
solar filaments. Hence, understanding the distribution and transport of  
magnetic flux at the solar surface is of primary importance to describe many aspects 
of solar magnetism.

To reproduce the surface distribution of magnetic flux, different types 
of processes are included in models: active regions dynamics, differential 
rotation, meridional circulation and diffusion across the supergranular 
network \cite[][]{Ossen03}. Supergranules are known to dominate the 
evolution and diffusion of surface magnetic fields underlying coronal and
heliospheric fields, as evidenced by the structure of both active and 
quiet magnetic networks. They are distributed 
over the entire solar surface, indicating that the corresponding 
velocity field is always present to carry magnetic flux. Moreover, 
evidence of connections between supergranulation, interplanetary magnetic 
field and solar energetic particules has been demonstrated with measurements  
of ions from the impulsive solar Flare with the Advanced Composition Explorer (ACE)
spacecraft and numerical simulations of the propagation of
energetic particles \cite[][]{GJM00}. Solar wind outflow
sources have also been located, with the SUMER/SOHO instrument,
along the boundaries and boundary intersections of the magnetic
network, which delineates supergranulation \cite[][]{HDLBC99}.
More recently,  \cite{AIOU08} reported evidence of continuous
reconnections at the boundaries of the supergranular network
lanes in the quiet Sun and coronal holes. He proposed
a scenario involving reconnections between the strong network
magnetic field at supergranule boundaries and the continuously
advected weak field from the supergranular cell interior.
From many points of view, it therefore seems essential to
have a good understanding of supergranular flows to 
construct accurate models of surface and atmospheric magnetic 
activity of the Sun.

Supergranulation is usually assumed to be the surface 
imprint of deeper convective motion. Why this scale is
so prominent amongst all the continuum of scales 
remains unexplained. As the convection zone
is strongly turbulent and highly stratified in density, numerical
modelling reaching supergranular scales has proven difficult
so far \cite[][]{DET04,RLR05,GZKBSN07} so that the dynamics at 
these scales remain poorly understood.
Supergranules are traditionally described as convection eddies
with horizontal flows diverging from a cell centre and subsiding
flows at the cell boundaries and horizontal currents associated
with each supergranule are believed to sweep magnetic
fields to its borders. Different theoretical approaches try to 
explain the driving of supergranulation \cite[see
a summary in][]{DET04}. For instance, the spatial correlations
between exploding granules may drive a large-scale instability
injecting energy at supergranular scales \cite[][]{RRMR00};
another scenario suggests that granules impose
fixed thermal flux boundary conditions, triggering a convection flow 
at scales larger than granulation \cite[][]{RR03b}.
\cite{RAST03} suggested that interacting downwards flows
cluster and produce the scale of meso or supergranules.
The picture is further complicated by the results of \cite{GDS03} and \cite{SCHOU03}
who suggested that supergranulation was associated
with a wave pattern;  \cite{RLT04} and \cite{LISL04},
using a similar approach, however argued that the spectrum 
of supergranulation was instead consistent with two nonoscillatory
flows identified with the mesogranular and supergranular scales
having different rotation rates. Overall, it is clear that the 
physical nature of supergranulation is still in debate and that new 
accurate measurements are required in order to obtain more clues
regarding its physical origin.  Addressing this problem most notably
requires high-resolution observations of the solar surface dynamics over a 
large field-of-view and for several consecutive hours.

In this paper, we study photospheric motion using new data obtained with 
the Hinode satellite, that fulfill some of these requirements. We particularly
focus on their relation to the magnetic field network, by identifying and 
following  Tree of Fragmenting Granules (TFGs) introduced by \cite{RLRBM03}.
These features, also called families of granules, are an
appropriate tool to quantify the temporal and spatial organization
of the solar granulation at larger scales. In Section 2 we describe
an exceptionally long (48h), multi-wavelength, time-sequence
of observations obtained with Hinode/SOT. New statistics on
TFGs are given in Section 3. Section 4 describes a correlation between
the photospheric network and the evolution of a passive
scalar and magnetic field location. Our conclusions are presented
in Section 5.

\begin{figure}
\centerline{\psfig{figure=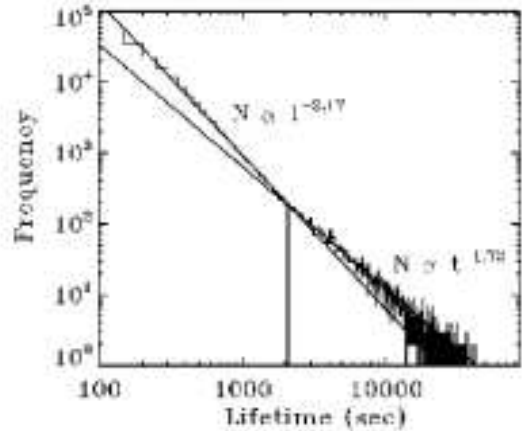,width=9 cm}}
\caption[]{Lifetime histogram of individual granules and TFGs
for the full time sequence of 48h.}
\label{powerlaws}
\end{figure}

\section{Observations and data reduction}

We used multi-wavelength data sets of the Solar Optical Telescope (SOT),
aboard the Hinode \footnote{The Hinode spacecraft, launched in 2006,
was designed and is now operated by JAXA (Japanese Space Authorities)
in cooperation with NASA (National Aeronautics and Space Administration)
and ESA (European Space Agency)}. mission \cite[e.g.][]{STISO08,ITSSO04}.
The SOT has a 50~cm primary mirror with a spatial resolution of about
0.2~\arcsec at 550~nm.

For our study, we used multi-wavelength  observations from Hinode/SOT NFI 
(NarrowBand Filter Imager) and BFI (Broadband Filter Imager). This instrument
measures the Stokes profiles of the FeI line at 630.2 nm with a spatial
resolution 0.16~\arcsec.  More precisely, the Lyot filter is set to a 
single wavelength in the blue wing of the line, typically $-120$ m\AA for the Fe I 
magnetograms. Then images are taken at various phases of the rotating polarization 
modulator and added or subtracted into a smart memory area in the on-board computer.
The units of both I and V images in both Level-0 and Level-1 are  "data numbers" or DN.
The  BFI scans consist of time sequences obtained in the Blue Continuum (450.4nm), 
G band (430.5nm) and  Ca II H (396.8nm).

The observations were recorded continuously from 29 August 10:17 UT
to 31 August 10:19UT 2007, except for a 7 minute interruption at the
disk center on 30 august at 10:43 UT. The solar rotation is compensated for 
in order to observe exactly the same region of the Sun. The time step
between two successive frames is 50.1 sec. The field of view with BFI
observations is $111\farcs6\times111\farcs6$ with a pixel of 0\farcs109
($1024\times1024$). After alignment, the useful field-of-view reduced
to $100~\arcsec\times92~\arcsec$.  

 To remove the effects of oscillations, we applied a subsonic Fourier filter. 
This filter is defined by a cone in the $k-\omega$  space, where k and $\omega$ are spatial
and temporal frequencies. All Fourier components such that $\omega/k\leq V=6\,\mathrm{km~s}^{-1}$ 
are retained so as to keep only convective motion  \cite[][]{TTTFS89}.
To detect TFGs, granules were labeled in time as described in \cite{RLRBM03}.

\begin{figure}
\centerline{\psfig{figure=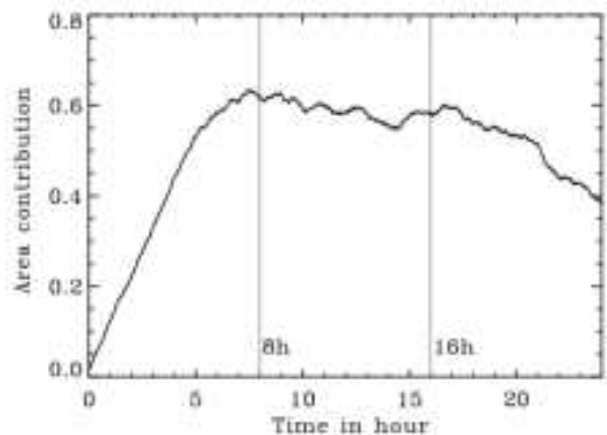,width=9 cm}}
\caption[]{Evolution of the percentage of granules belonging to TFGs
existing longer than 8h.}
\label{context2b}
\end{figure}

\begin{figure}
\centerline{\psfig{figure=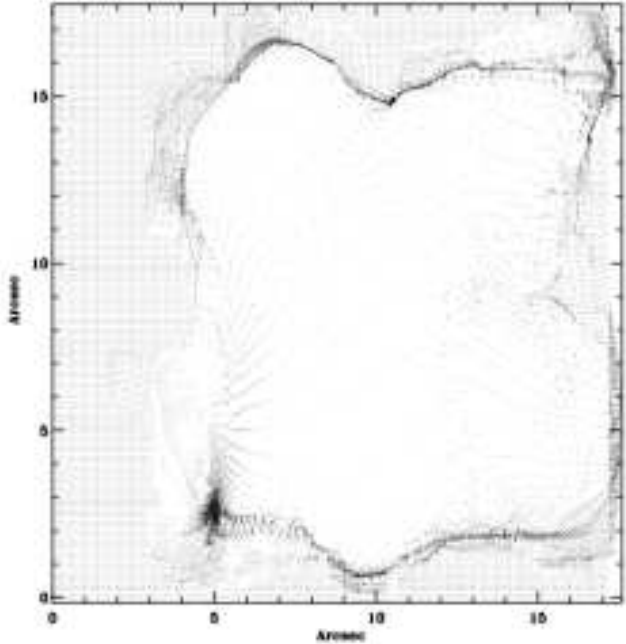,width=9 cm}}
\caption[]{Expulsion of corks of the TFG 2857 in 
approximatively 2 hours. The box has the size of the mesogranule scale.}
\label{context8}
\end{figure}

\section{TFG detection and temporal properties}

\begin{figure*}[t]
\centerline{\psfig{figure=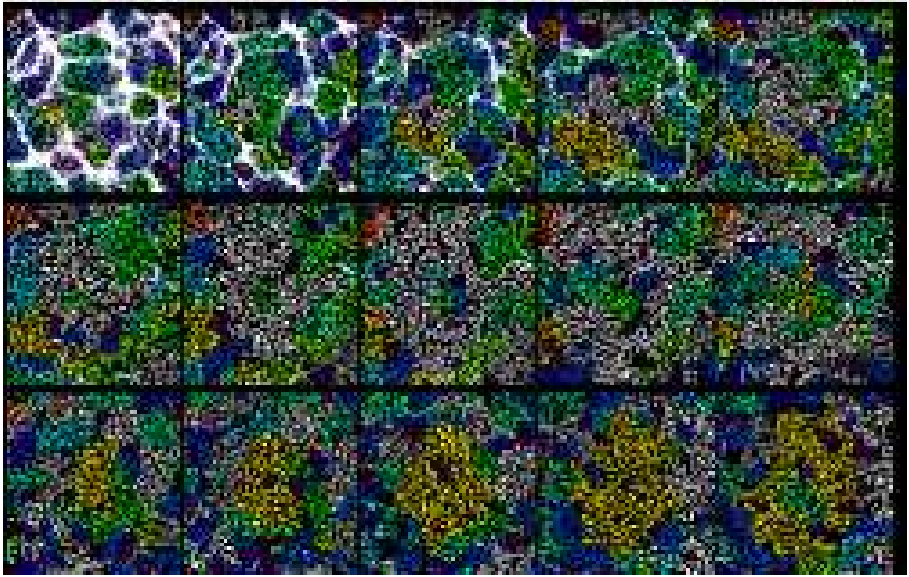,width=12cm}}
\caption[]{Example of families evolution over 15~hours with a time step of 1 hour.
The field of view is $38.2~\arcsec\times39.9~\arcsec$. White indicates a TFG 
lifetime of 24h, although yellow indicates a TFG lifetime of 20h. Green and blue 
stand for TFGs with a lifetime shorter than 18h.   The families are colored (online)
according to their lifetime: 
white (24h and more), red (23 to 20 h), orange (19 to 16h), yellow (15 to 12h), 
green (11 to 8h), blue (7 to 4h) and purple (3 to 1h).}
\label{cork_fam} 
\end{figure*}

\begin{figure*}
\centerline{\psfig{figure=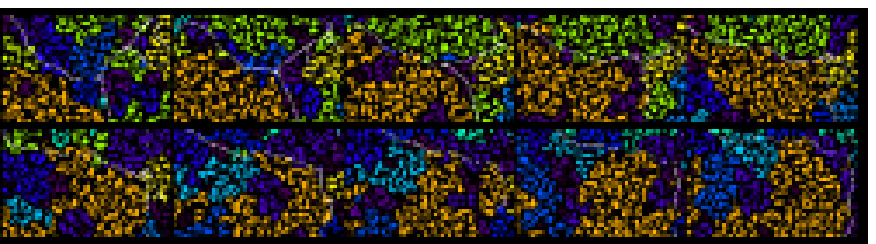,width=12cm}}
\centerline{\psfig{figure=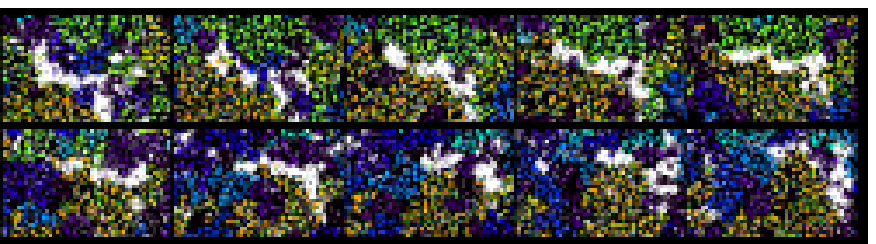,width=12cm}}
\centerline{\psfig{figure=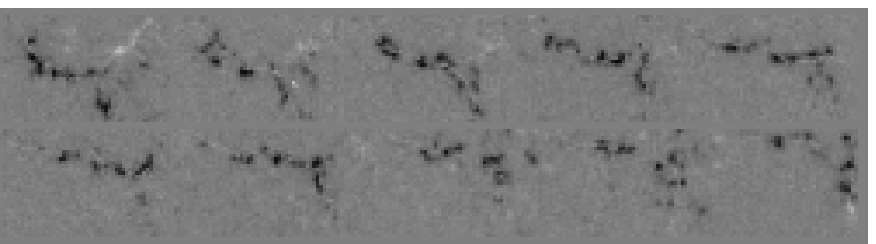,width=12cm}}
\caption[]{Corks (top) and magnetic field modulus (middle) location with respect
to families during 10h, $\delta t=1$h. Corks and magnetic field (Stokes V image) relative
location (bottom). The field of view is $24~\arcsec\times15.3~\arcsec$.}
\label{magnified}
\end{figure*}

3D analysis $(x,y,t)$ of the granular intensity field demonstrates that
a significant fraction of the granules in the photosphere are organized
in the form of TFGs \cite[][]{RLRBM03}.  A TFG consists of a family of
repeatedly splitting granules, originating from a single granule.
Since our data set is composed of two continuous sequences of 24h
separated by a gap of 7 minutes, we first analyse separately each of the
24h series and then connect them together to determine the evolution
of TFGs over 48h.

As shown in Fig.~\ref{powerlaws}, the distribution of the lifetime of
the TFGs can be fitted by two power laws with exponents $-2.17$ and
$-1.72$ for TFGs shorter and longer than 2100~s respectively. The second
 exponent, $-1.72$, is identical to the one previously found  with a 8.75 h 
image series obtained with the Swedish Vacuum Solar Telescope at La Palma 
\cite[][]{RLRBM03}. Howewer, we find a different exponent
for TFGs with lifetimes shorter than 2100$\pm100\,\mathrm{s}$. This time
scale seems to correspond to the longest isolated granules, which do
not create a TFG. The Hinode data which are free of atmospheric seeing 
allows us to continuously  follow the smallest granules. This explains why 
we detect more TFGs with shorter lifetimes than with the ground based La Palma 
data. We conclude on the existence of two different kinds of TFGs characterized 
by their lifetime.  The shorter lifetime TFGs appear scattered everywhere
in the field of view between the longest TFGs.  The slope of $-1.72$ for
the longest TFGs is robust and easily reproducible; 
 It would be interesting in the future to determine this slope using 
large cartesian numerical simulations of solar surface convection including 
scales up to the supergranulation scale \cite[e. g.][]{GZKBSN07}.

As some TFGs turn out to last longer than 24h, we connected the two time series,
filling the 7~min gap with the
following strategy: using the first time sequence, we compare the distribution
of families at $t=$24h-7mn and at $t=$24h. We observe that the area variation was
not more than 3\%. We thus assumed that during the gap, the change was not
larger and decided that families at $t=$24h could be reported as such at
$t=$24h+7mn, i.e. at the beginning of the second series. The alignment of the two
sequences was also checked using the CaII and G-band data. In doing so, we managed
to make an image sequence lasting 47h23min and covering a field of view of
$89~\arcsec \times 103~\arcsec$. Using this very long sequence, we could improve 
the statistics of data in Fig.~\ref{powerlaws}, and found that the lifetime of TFGs 
could reach 43h50mn (within our field).

Although the number of TFGs decreases with lifetime, long-lived TFGs
cover a significant fraction of the solar surface. This is shown in
Fig.~\ref{context2b}, where the fraction of the total area covered by
granules belonging to TFGs living more than 8h is displayed using the
first 24h-sequence. This plot shows that 60\% of the area is covered
by families living longer than 8h. This fraction steadily increases 
during the first 8 hours because, during this period of time, many granules 
are still associated with families that were born before the beginning of the
time-sequence and therefore appear to have a shorter lifetime.
It also slightly decreases at the end of the sequence 
since the time window cuts the last born TFGs.  If we now include more short-lived 
TFGs, reducing the lifetime threshold to 2 hours, then we find that the surface 
coverage increases up to 88\%. This is greater than the 62\% that we found
with the La Palma data. The difference is due to a better labeling
of granules (no seeing and distortion), and a longer duration of the
sequence allowing better statistics.

\section{Network and cork distributions}

\subsection{Velocity fields}

To further investigate the solar surface dynamics, we now
focus our attention on the velocity fields and their transport properties. The
velocity field is determined using the classical local correlation
tracking (LCT) algorithm \cite[e.g.][]{RRMV99} using a spatial window
FWHM of 3~\arcsec and a temporal window of 30 minutes. The typical rms
velocity is 300~m/s. As in \cite{MTRR07} and \cite{RMRRBP08}, we find that the 
probability distribution functions of divergence at scales larger than 10~Mm show 
a net excess, compared to a Gaussian distribution, in the positive wing, suggesting 
intermittency in solar surface convection.

\subsection{Cork motions}

\begin{figure}
\centerline{\psfig{figure=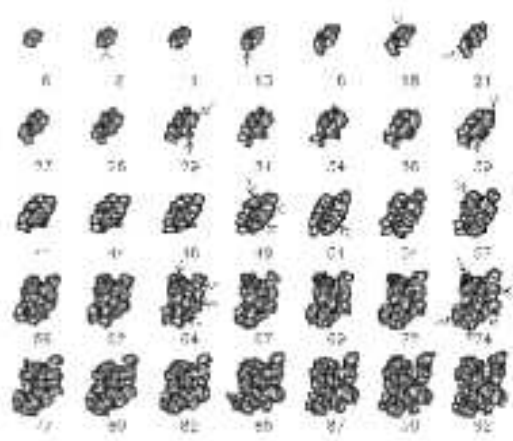,width=9 cm}}
\caption[]{Evolution of a TFG: time is given in minutes and arrows
indicate the location of the granule explosion.}
\label{context10}
\end{figure}

\begin{figure}
\centerline{\psfig{figure=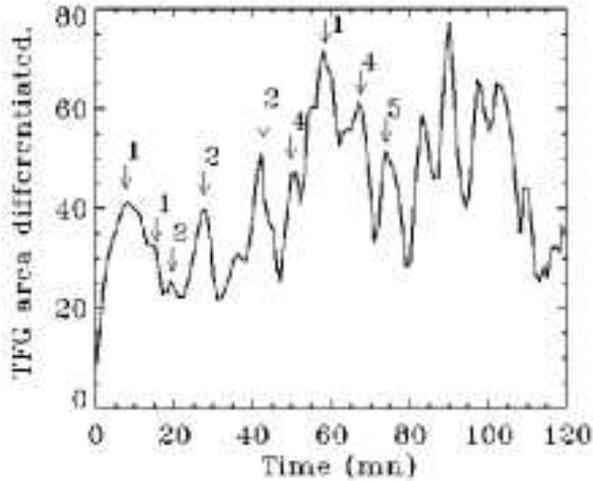,width=9 cm}}
\caption[]{Time evolution of the derivative of the area of the TFG shown in
Fig.~\ref{context10}. Numbers refer to the number of explosions indicated 
by arrows in Fig.~\ref{context10}.}
\label{context9}
\end{figure}

The transport properties at the surface of the turbulent velocity field are best
 illustrated using the evolution of an initially uniformly distributed
passive scalar; the image is classically that of floating corks
\cite[][]{SBN94}. The corks are deposited on the meshes of a regular grid
at initial time $t=0$ and their surface distribution is followed during the
whole sequence as they are advected by the  horizontal velocity field.
 Fig.~\ref{context8} displays the corks distribution resulting from 
the evolution of a TFG depicted in Fig.~\ref{context10}. The isolated action of the 
velocity field within an isolated TFG leads to the advection of 90\% corks at a 
mesogranular scale on a timescale of between two and four hours (flows are not taken
into account outside of the isolated TFG). 
A comparison with the underlying families shows that they are simply expelled from 
a given family. The first snapshots of Fig.~\ref{cork_fam}
clearly show this phenomenon. At later times, the corks remain more or less at 
the TFG boundaries as can be seen in Fig.~\ref{magnified}.
 As we have shown in our previous paper \cite[][]{RLRBM03} (fig,11,12 and 13),  
TFGs are equivalent to a diverging flow at mesoscales. 
So, this scale is persistent up to 43h corresponding, in the  Hinode data, to the
TFG itself or to some TFG branches for the longest living TFG. The large scale 
TFGs are probably the smallest supergranular sizes.

Investigating the mechanism of this fast
transport, we  found that exploding granules are the key features
generating these motions. Fig.~\ref{context10} displays in detail the
evolution of a family; the arrows indicate the
location of the observed explosion during 1h25mn. The evolution
of the time derivative of the family area is plotted in Fig.~\ref{context9}. It
demonstrates that each increase of the total area of this family
is related to granule explosions whose number is quoted over each peak
of the curve.  Hence, we observe that the family grows in area by
a succession of granule explosions, which are correlated in time,
thereby expelling the corks.
\begin{figure}[t]
\centerline{\psfig{figure=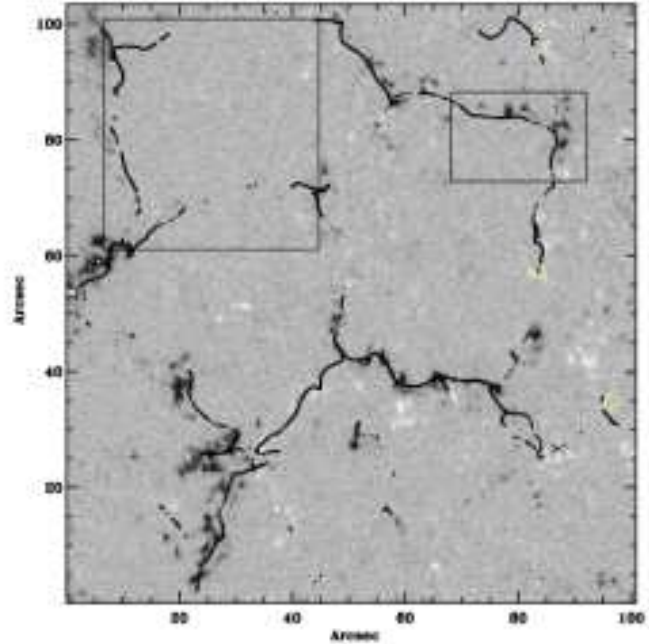,width=9 cm}}
\caption[]{ Cork location at the end of the first time sequence of 24h relatively to 
the longitudinal magnetic field (Stokes V image). The left box shows the  field of view 
of Fig.~\ref{cork_fam} and the right box indicates the field of view of Fig.~\ref{magnified}}
\label{context3}
\end{figure}

\begin{figure}[t]
\centerline{\psfig{figure=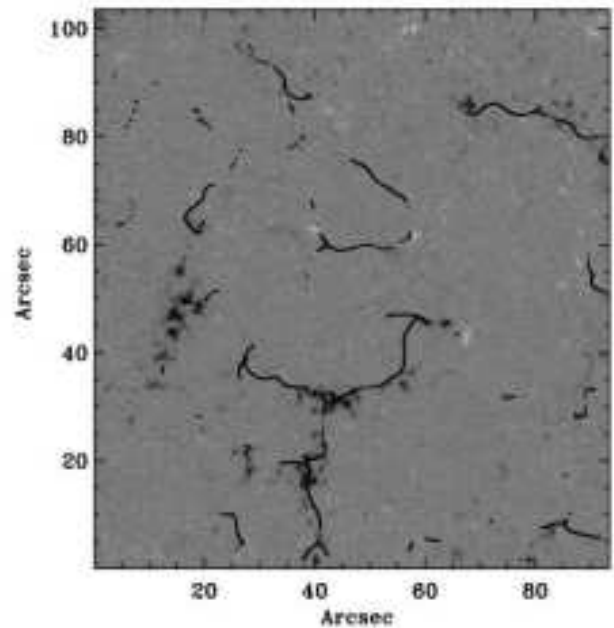,width=9 cm}}
\caption[]{Cork location at the end of the second time sequence of 24h
relative to the longitudinal magnetic field (Stokes V image).}
\label{context4}
\end{figure}

\begin{figure}[t]
\centerline{\psfig{figure=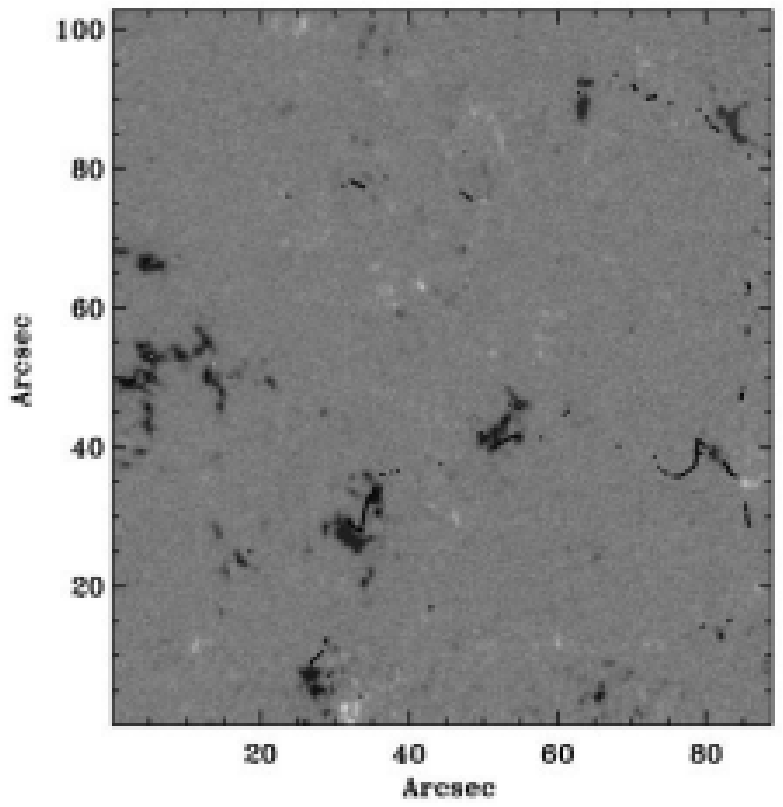,width=9 cm}}
\caption[]{Cork location at time $t=$36h relative to the longitudinal magnetic
field (Stokes V image).}
\label{cork_mag36}
\end{figure}

\begin{figure}[t]
\centerline{\psfig{figure=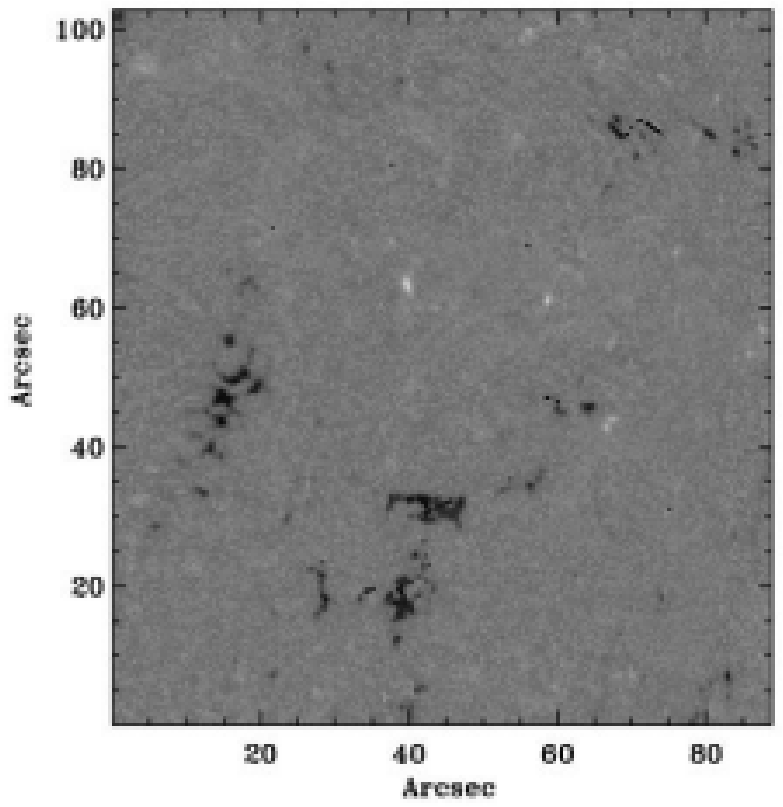,width=9 cm}}
\caption[]{Cork location at time $t=$47h relative to the longitudinal magnetic
field (Stokes V image).}
\label{cork_mag47}
\end{figure}

Fig.~\ref{cork_fam} also shows the connection of families and the
supergranular scale. In the second frame of this figure we note the 
birth of a white family. Then, the successive frames show its evolution
and its surface growing up to frame 9. In frame 10, one can see a
new yellow TFG birth, which growth in the following frames and pushes
the initial white TFG. 
 We observe that the distribution of corks at the supergranular scale 
also results from the combined action of long-lived TFGs that are close spatially 
and more or less in the same phase of their evolution. Their mutual interactions 
tend to expulse corks of the area they cover on 3h-4h and concentrate the corks 
in the magnetic network, at supergranular scales in approximatively 6h-8h. 
One example of this evolution can be seen in our data (and movies) in the 
center of the field of view where two long-lived TFGs are competing for their 
development. 
 
 On longer timescales, long living families tend to structure the  cork 
distribution either by successive birth at the same location, or by interaction
of two or more competing TFGs  which are located in close proximity  and 
in the same phase of their evolution. Overall, the formation of this pattern depends 
on the TFG relative size and lifetime but seems to be mostly governed by long-lived 
TFGs.

\subsection{Corks and magnetic fields}

When longer time scales are considered it is well known
that corks, like the magnetic field, concentrate on the supergranule
boundaries \cite[e.g.][]{RRMR00, KR03}.
 The quality of the data (free of atmospheric seeing) and the length of the time 
sequence allow us to provide a very clear illustration of this phenomenon.
 To determine the magnetic field distribution, we use Stokes V which is usually considered 
as a proxy for the radial magnetic field at the disk centers.

 In Fig.~\ref{magnified}, we see that
magnetic fields, like corks, lie at the boundaries of granule families. In 
this figure, we clearly observe that the spatial evolution of the magnetic 
field (or corks) are related to the geometric evolution of the long lived TFG 
(shown here in orange). This demonstrates the physical action of the TFG on 
the magnetic field patch displacement over the solar surface and shows that
plasma governs magnetic elements at the TFG scale in the quiet Sun.
Fig.~\ref{context3} and \ref{context4} illustrate the same result at a larger
scale when cork motion has been followed for 24h. Most of the corks
(70\%) are still in the field-of-view at the end of each sequence. They
exactly match the position of the highest magnetic flux concentration of the network,
which lies on supergranule boundaries. The network boundaries defined 
by these patches are incomplete, with spatially intermittent sites of high magnetic 
flux density well correlated with sites of strong convergence.

If we further monitor the motion of the corks using the 48h-sequence,
we find that after such a long time, 60\% of the corks remain in
the field-of-view, trapped in a small area of $\sim$10~Mm$^2$, which
also displays the most intense magnetic fields. This is illustrated in
Figs.~\ref{cork_mag36} and \ref{cork_mag47}. 

\begin{figure}[t]
\centerline{\psfig{figure=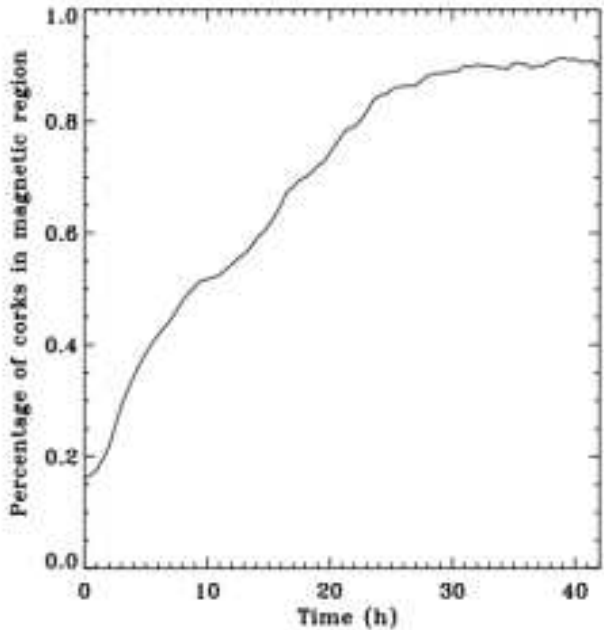,width=9 cm}}
\caption[]{Time evolution of the percent of corks  located in the region where the 
longitudinal magnetic field magnitude is greater than 248 Gauss, which corresponds to 
of magnetic patches at the supergranule limits.}
\label{context101}
\end{figure}

Furthermore, if we monitor the number
of corks inside magnetic regions (we use a threshold of  45 SOT data number to
detect the magnetic network, which corresponds, by using a raw calibration described 
in \cite[][]{CHAE07},  to 248 Gauss), we find, as shown in Fig.~\ref{context101}, that 90\%
of the corks in the field end up in magnetic patches. Furthermore, monitoring the number 
of corks inside magnetic regions, we find that 90\% of the corks in the field end up in 
magnetic patches (see Fig.~\ref{context101}). Note that a threshold of 45 SOT data number 
is used in Stokes V images to detect the magnetic network. Using a raw calibration  
described in Chae et al. 2007, this number correspondsto a field amplitude of 248 Gauss).
 This is particularly visible
in movies\footnote{\tiny\texttt{http://www.lesia.obspm.fr/\~{}malherbe/papers/Hinode2008/}}
of the evolution of both the magnetic field and the TFGs, which show that
magnetic fields, squeezed between TFGs, follow their displacement while drifting
along their boundaries; at some time, they meet a vertex of TFG boundaries and become 
stabilized there, forming a patch.

 As noted before, this patchy distribution of magnetic fields is closely associated 
with sites of strong flow convergence. These sites coincide with vertices of supergranular 
flow cells and presumably correspond to downflows, as required by mass conservation on the 
supergranular scale  \cite[][]{CCT07}. The observed stabilization of magnetic 
patches at supergranular scales possibly indicates that the dynamical Lorentz force
feedback provided by these concentrated patches is sufficiently strong to oppose 
the flow at scales comparable to the supergranular scale, which might explain
the emergence of this particular scale in the velocity power spectrum  of the quiet Sun.

\subsection{Horizontal turbulent diffusion}

Lastly, cork motion allows the measurement of the horizontal turbulent
diffusivity at the surface of the Sun. The turbulent magnetic diffusion is an
important quantity for mean-field dynamos  \cite[][]{BS05,RE08} as it controls 
the timescale of the dynamo oscillations (in the case of a time-periodic dynamo). 
Thus several authors have evaluated this quantity. For instance, \cite{STW95}, using a
kinematic model tuned to match supergranulation properties, find diffusivities
in the range 500 to 700~km$^2$/s. With the very long
time series that we have at our disposal, we can make a direct evaluation of
this coefficient based on the precise plasma flow. However, because we use
granules to determine the flow, the horizontal velocity fields are filtered
at all scales below 2.5Mm \cite[see][]{RRLNS01}. Thus we give an estimate of
the diffusivity imposed by the scales above 2.5Mm.

\begin{figure}
\centerline{\psfig{figure=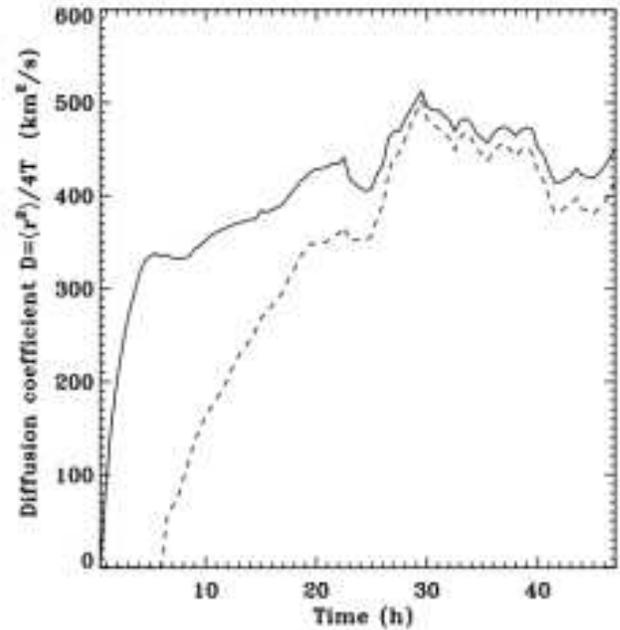,width=9 cm}}
\caption[]{Diffusion coefficient over 48h (solid line) and the diffusion coefficient computed from 6h to 48h
with a starting reference position at time 6h, when the corks are at the mesoscale border (dashed line)}
\label{context11}
\end{figure}

As in other works, the diffusion coefficient is estimated from $D=\langle
r^2\rangle/4t$, where $\langle r^2\rangle$ means the mean displacement
of all particles (corks) at time $t$. As shown in Fig.~\ref{context11}
(solid line), after a rapid growth due to advection by the dominant
mean flow (typically the mesogranular scale velocities), the diffusion
 coefficient seems to converge to a value of $430\,\mathrm{km}^2\,\mathrm{s}^{-1}$.
 The diffusion coefficient also was computed with a starting
reference position at time 6h where the corks are at the mesoscale
boarder. We observe that the cork motion along the mesoscale converge at 
around 29 hours to the same value of the diffusion coefficient near
a value of 430~km$^2$/s. This indicates that corks and patches
of corks are always in movement due to the TFG evolution. The
figure shows, for the first time, the importance of using long time
series for  this estimate. Saturation is indeed obtained only after
24hrs. With shorter time sequences we  could expect a growth beyond
the present value. Finally, even though this may be a coincidence, 
we note that the implied diffusion timescale over the solar magnetic 
belts, namely ($\lambda R_\odot$)$^2$/430, is 11yrs ($\lambda\simeq32°$).

\section {Discussion and conclusion}

Thanks to the Solar Optical Telescope onboard  Hinode, we have been
able to monitor with high spatial and temporal resolution, free
of atmospheric seeing, a region of the Sun's photosphere for two
days. The field was wide enough to contain 4 supergranules and thus to 
examine their internal dynamics.

Using the TFG concept we could quantify the temporal and spatial
correlations in the evolution of granules. The typical size of
mature TFGs suggest that they trace some mesoscale rising plumes. In
 the previuos debate on the specificity of mesogranulation as a genuine
scale of sub-photospheric convection rather then a mere extension of
granular flow \cite[][]{SB97,RRMR00}, TFGs point towards the genuine
scale. In this scenario, granulation would be the boundary layer of these
flows. However, since solar convection is driven by the strong cooling
of the Sun's surface, it may still be that mesoscale flows are forced
by the granulation layer. At this stage, it is clear that numerical
simulations are necessary to better understand the connection between
the dynamics of the surface layers and that of the deeper ones.

From our analysis, we observe that most of the granules visible on the 
solar surface belong to the long-lived TFGs, indicating that these families 
structure the flow. The evolution of floating corks demonstrates that the final 
state of passively advected quantities like weak magnetic elements is a patchy 
distribution on the boundaries of a supergranule, which correspond presumably
to stable downward flows.  The stabilization of magnetic patches 
in these areas suggests that supergranulation could be an emergent 
length scale building up as weak small-scale magnetic elements are advected and 
concentrated by TFG flows, occasionally colliding and aggregating to form larger
and stronger magnetic patches, eventually resisting the flow dynamically. 
The temporal evolution  of magnetic elements, corks and  TFGs are  available  
at {\small \texttt{http://www.lesia.obspm.fr/\~{}malherbe/papers/Hinode2008/}}.
  
Two kind of interactions between TFGs are mainly observed to diffuse
the corks to the limit of supergranular scales. The first one corresponds to
successive long-lived TFG births at the same location, whose combined motions
sweep corks to network scales; parts of older TFGs, with a classical mesoscale,
are pushed away by new TFGs, thereby contributing to the formation of a larger 
scale comparable to the supergranular scale. The second process is the common action 
of two or more close long-lived TFGs born more or less at the same time, which push
corks to the network limit. The temporal co-spatiality between magnetic fields
and corks in our time sequence analysis implies that the evolution of
the distribution of magnetic fields is clearly driven by TFGs on the
Sun's surface. A relation between mesoscale flows and the distribution
of magnetic elements was also reported by \cite{DC03}, \cite{DCSA03}, \cite{rouppe05} 
and  \cite{LSKB07} .

Finally, the length of the monitoring time window  (48h) allowed us
to directly estimate the diffusion coefficient induced by horizontal
motion at scales larger than 2.5~Mm. We found a value of $430\,\mathrm{km}^2\,\mathrm{s}^{-1}$,
which might be helpful to design mean-field dynamo models.

 Satellites like Hinode/SOT or the future SDO as well as 
large field-of-view ground based instruments like CALAS at Pic du Midi
now enable us to study the full complexity of the interactions of multiple scales of 
convection and magneticfields at the solar surface. In this paper
we have shown that identifying and following coherent structures like TFGs 
can help us understand the detailed physics of these interactions and how
they relate to the solar cycle.

\begin{acknowledgements}

We are grateful to the Hinode team for the possibility to use their data.
Hinode is a Japanese mission developed and launched by ISAS/JAXA,
collaborating with NAOJ as a domestic partner, NASA and STFC (UK)
as international partners. Scientific operation of the Hinode mission
is conducted by the Hinode science team organized at ISAS/JAXA. This
team mainly consists of scientists from institutes in the partner
countries. Support for the post-launch operation is provided by JAXA
and NAOJ (Japan), STFC (U.K.), NASA, ESA, and NSC (Norway).This work
was (partly) carried out at the NAOJ Hinode Science Center, which
is supported by the Grant-in-Aid for Creative Scientific Research
"The Basic Study of Space Weather Prediction" from MEXT, Japan (Head
Investigator: K. Shibata), generous donations from Sun Microsystems,
and NAOJ internal funding.
We thank the CNRS/INSU for support of the one-year sabbatical of 
D. Brito at LATT (while member of LGIT, Universit\'e Joseph-Fourier, 
CNRS in Grenoble).
This work was supported by the Centre National de la Recherche
Scientifique (C.N.R.S., UMR 5572), by the Programme National Soleil Terre
(P.N.S.T.).

\end{acknowledgements}

\bibliographystyle{aa}
\bibliography{biblio}

\end{document}